\documentclass[a4paper,twocolumn]{article}
\usepackage{graphicx}

\def\be{\begin{equation}}
\def\ee{\end{equation}}
\def\bea{\begin{eqnarray}}

\def\eea{\end{eqnarray}}

\begin{document}

\title{BEC Collapse, Particle Production and Squeezing of the Vacuum}
\author{E. A. Calzetta$^1$ 
and B. L. Hu$^2$  \thanks{Emails: calzetta@df.uba.ar,
hub@physics.umd.edu}\\ $^1${\small Departamento de Fisica, FCEyN
Universidad de Buenos
Aires Ciudad Universitaria, 1428 Buenos Aires, Argentina}\\
$^2${\small Department of Physics, University of Maryland, College Park, MD
20742, USA}}
\date{{\small (Second revision June 18, 2003; First revision Jan 18, 2003.
Original version Aug 28, 2002. umdpp 03-006)}} \maketitle

{\it Phenomena associated with the controlled collapse of a
Bose-Einstein condensate described in the experiment of Donley et
al \cite{JILA01b} are explained here as a consequence of the
squeezing and amplification of quantum fluctuations above the
condensate by the condensate dynamics. In analyzing the changing
amplitude and particle contents of these excitations, our simple
physical picture provides excellent quantitative fits with
experimental data on the scaling behavior of the collapse time
and the amount of particles emitted in the jets.}

\bigskip In the experiment described by Donley et al. \cite{JILA01b}, a
Bose-Einstein condensate (BEC) 
in a cold ($3$nK) gas of Rubidium atoms is rendered unstable by a
sudden inversion of the sign of the interaction between atoms.
This is done by altering the binding energy at Feshbach resonance
with an external magnetic field. After a waiting time
$t_{collapse},$ the condensate implodes (called Bose-Nova), and a
fraction of the condensate atoms are seen to oscillate within the
magnetic trap which contains the gas. These atoms are said to
belong to a `burst'. After a time $\tau _{evolve}$
the interaction is suddenly turned off. For a certain range of values of $%
\tau _{evolve},$ new emissions of atoms from the condensate are
observed. They are called `jets'. Jets are distinct from bursts:
they are colder, weaker, and have a characteristic disk-like
shape. \footnote{ We call attention to the distinction between the
'Bose-Nova' \cite{JILA01b} experiment studied here and other BEC
collapse experiments \cite{JILA98,JILA02a}. At magnetic fields
around
$160$G, where the effective scattering length is of the order of $500a_{0}$ (and positive)($%
a_{0}=0.529$\ $10^{-10}$m\ \ \ is the Bohr radius) it is possible
to observe oscillations between the usual atomic condensate and
the molecular state \cite{JILA02a} with a frequency of
oscillations of hundreds of KHz \cite{KGB02}. By contrast, in the
'Bose-Nova' experiment \cite{JILA01b} typical fields were around
$167$G, the scattering length was only tens of Bohr radii (and
negative) and the frequency of atom - molecule oscillations may
be estimated as well over ten MHz \cite{JILA03}. While coherent
resonance between the atoms and the molecules is expected to
exist for all of these experiments,  and has been shown to play
an important role in the outcomes of some \cite{JILA03}, we deem
it unlikely that it plays a dominant role in this experiment
other than renormalizing the scattering length (For details, see
\cite{CHBosenova}). Indeed no oscillations are reported in the
original experimental paper. Instead, as this letter shows, the
primary mechanism for the Bose Nova phenomena is the parametric
amplification of quantum fluctuations by the condensate dynamics,
resulting in bursts and jets as particle production from (the
squeezing of) the vacuum. Recent numerical simulations
\cite{Savage} and rigorous theoretical investigations \cite{DS02}
indicating the inadequacy of mean field theory seem to
corroborate this view.}

The model is based on the Hamiltonian operator for $N$ interacting
atoms with mass $M$ in a trap potential $V\left( \mathbf{r}\right)
=(\omega _{z}^{2}z^{2}+\omega _{\rho }^{2}\rho ^{2})/2$,
with radial $\rho $ and longitudinal $z$ coordinates measured in units \footnote{%
We use a sign convention such that the effective coupling
constant is positive for an attractive interaction,  and a system
of units where the length $a_{ho}$, time $t_{ho}$ and energy scale
$E_{ho}=\hbar \omega =M\omega ^{2}a_{ho}^{2}$ are defined with
reference to the average frequency $\omega $. We work with units
such that these three scales take the value $1$.}. of $%
a_{ho,}$ where $a_{ho}$ is a characteristic length of the trap,
with associated (dimensionless) frequencies $\omega _{z}=\omega
_{axial}/\omega \sim 1/2$ and $\omega _{\rho }=\omega
_{radial}/\omega \sim \sqrt{2}.$ The interaction is assumed to be
short ranged. We introduce a dimensionless field operator
$\mathbf{\Psi }\left( r\right) \equiv a_{ho}^{-3/2}\Psi \left(
x\right) $,  and a dimensionless coupling constant $u=\left( \hbar
\omega a_{ho}^{3}\right) ^{-1}U=4\pi \left( a/a_{ho}\right) $.

$\Psi $ obeys the equation of motion
$\dot{\Psi}=i\left[\hat{H},\Psi \right] $ and satisfies the equal
time commutation relations $\left[ \Psi \left(
t,\mathbf{r}\right) ,
\Psi ^{\dagger }\left( t,\mathbf{r^{\prime }}%
\right) \right] =\delta ^{\left( 3\right) }\left( \mathbf{r}-\mathbf{%
r^{\prime }}\right) .$ We decompose the Heisenberg operator $\Psi =\Phi (%
\mathbf{r},t)+\psi (\mathbf{r},t)$ into a c-number condensate amplitude $%
\Phi $ and a q-number noncondensate amplitude $\psi $, consisting
of the fluctuations or excitations. We obtain the equation of
motion for the fluctuation field by subtracting from the full
Heisenberg equation the Gross - Pitaievsky equation (GPE)
governing its own expectation value under the self-consistent
mean field approximation, $\psi ^{\dagger }\psi \simeq
\left\langle \psi ^{\dagger }\psi \right\rangle ={\tilde{n}}$,
$\psi ^{2}\simeq \left\langle \psi ^{2}\right\rangle
={\tilde{m}}$ and $\psi
^{\dagger }\psi ^{2}\simeq 0$. We next parametrize the wave functions as $%
\Phi =\Phi _{0}e^{-i\Theta },$ $\psi =\psi _{0}e^{-i\Theta }$,
where $\Phi _{0}$ and $\Theta $ are real. During the early stages
of evolution, we may regard the condensate density as time
independent, and the condensate phase as homogeneous, $\Phi
_{0}=\Phi _{0}\left( r\right) ,$ $\Theta =\Theta \left( t\right)
$. We may then write the equation for the fluctuation field
\begin{equation}
\left[ i\frac \partial {\partial t}-H+E_0\right] \psi _0+u\Phi _0^2\left(
\psi _0+\psi _0^{\dagger }\right) =0
\end{equation}
where $E_{0}=\frac{1}{2}\left( \omega _{z}+2\omega _{\rho }\right)
$. To solve
this equation\footnote{%
The squeezing of quantum unstable modes and its back reactions on
the condensate has been considered before, e.g., as a damping
mechanism for coherent condensate oscillations \cite{KM01}, but
the application to the description of condensate collapse has up
to now been mostly qualitative \cite {Y02}.} we decompose it into
a self-adjoint and an anti-adjoint part $\psi _{0}=\xi +i\eta $,
each part satisfying an equation
\begin{equation}
\frac{\partial \xi }{\partial t}=\left[ H-E_{0}\right] \eta
\label{treintaydos}
\end{equation}

\begin{equation}
\frac{\partial \eta }{\partial t}+\left[ H-E_0-2u\Phi _0^2\right] \xi =0.
\label{treintaytres}
\end{equation}

Since the trap Hamiltonian is time - independent, we have

\begin{equation}
\frac{\partial ^2\xi }{\partial t^2}+\left[ H-E_0\right] H_{eff}\xi =0.
\label{treintaycuatro}
\end{equation}
Here $H_{eff}=H-E_{0}-2u\Phi _{0}^{2}$. To have an unstable
condensate it is necessary that at least one of the eigenvalues
of $H_{eff}$ is negative; the boundary of stability occurs when
the lowest eigenvalue is exactly zero. One further consideration
is that we are interested in the part of the fluctuation field
which remains orthogonal to the condensate, since fluctuations
along the condensate mode may be interpreted as condensate
fluctuations rather than particle loss \cite{M99}. The ground state of $%
H_{eff}$ is certainly not orthogonal to the condensate, since
neither have nodes.

If we adopt the values $\omega _{z}=1/2,$ $\omega _{\rho
}=\sqrt{2},$ relevant to the JILA experiment, then instability
occurs when $\kappa =N_{0}a_{crit}/a_{ho}=0.51$. This result
compares remarkably well with the experimental value $\kappa
=0.55$ \cite{JILA01b,JILA03}, as well as with the theoretical
estimate presented in Ref. \cite{GTT01}. This agreement may be
seen as natural, as the equations we postulate for the
fluctuations may be obtained from the
linearization of the GPE, discarding both ${\tilde{n}}$ and ${%
\tilde{m}}$. In both calculations, the geometry of the trap plays a
fundamental role.

\paragraph{Scaling of $t_{collapse}$ and Critical Dynamics}

As we have already noted, even for condensate densities above the
stability limit, no particles are seen to be lost from the
condensate during a waiting time $t_{collapse}.$ Experimentally,
$t_{collapse}$ is seen to get very large when  the threshold of
stability is  approached from above, in a way which closely
resembles the \textit{critical slowing down} near the transition
point characteristic of critical dynamics. In our problem, the
quantity which plays the role of relaxation time is the
characteristic time $\varepsilon ^{-1}$ of exponential growth for
the first unstable mode. This quantity diverges at the stability
threshold, which in our analogy corresponds to the critical
point. By dimensional analysis, we are led to the estimate
$t_{collapse}\sim \varepsilon ^{-1}.$ Close to the critical
point, we find

\begin{equation}
t_{collapse}=t_{crit}\left( \frac{a}{a_{cr}}-1\right) ^{-1/2}
\label{scaling}
\end{equation}

The power law Eq. (\ref{scaling}) describes with great accuracy the way $%
t_{collapse}$ scales with the scattering length; the best fit to the
experimental data is obtained for $t_{crit}\sim 5$ms.

\begin{figure}[h]
\includegraphics[height=4cm]{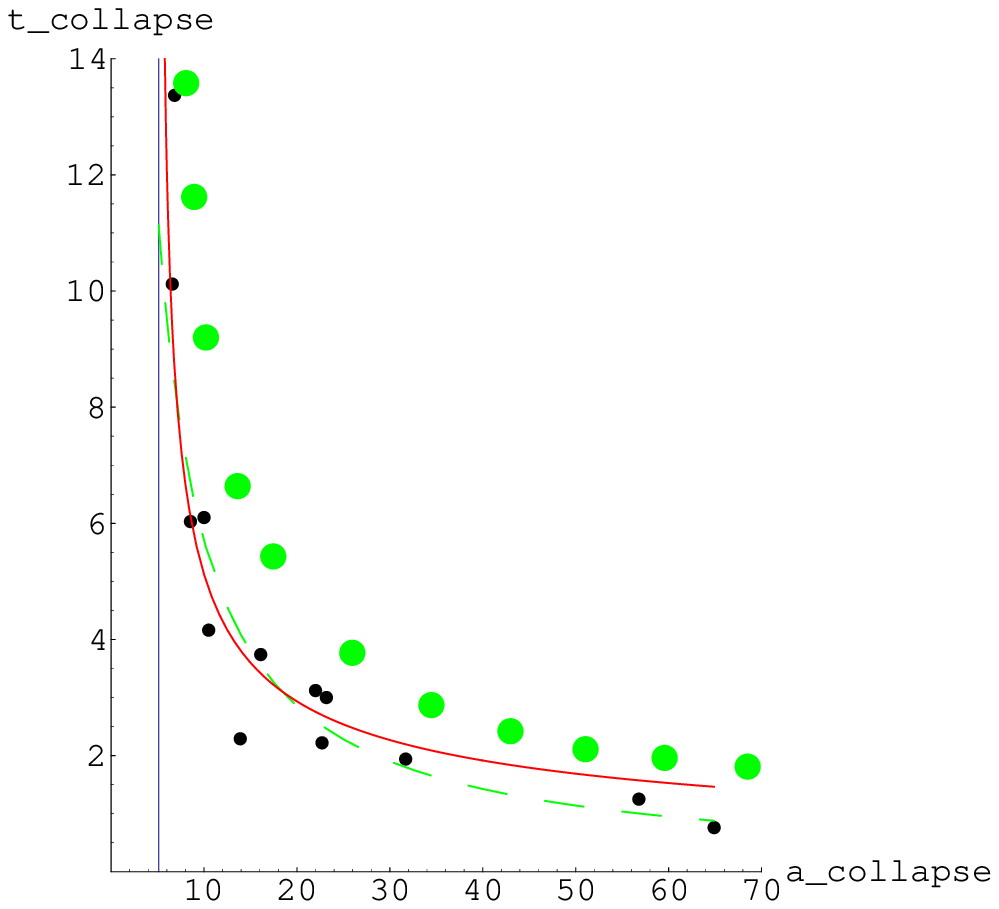}
\caption{}
\end{figure}

In Fig. 1 we plot the scaling law (\ref{scaling}) (full line)
derived here and compare it with the experimental data for
$N_{0}=6000$ as reported in Refs. \cite{JILA01b} (small black
points), the $t_{NL}\sim \left( uN_{0}\right) ^{-1}$ prediction
(suitably scaled) as given in \cite {Y02,TBJ00} (dashed line) and
the results of numerical simulations reported in \cite{SU03}
(large grey dots). While all three theoretical predictions may be
considered satisfactory, the $t_{NL}\sim \left( uN_{0}\right)
^{-1}$ behavior fails to describe the divergence of $t_{collapse}$
as the critical point is approached, and the results of numerical
simulations reported in \cite{SU03} based on an improved
Gross-Pitaevskii equation tend to be systematically above the
experimental results, which may be a further indication of the
quantum origin of this phenomenon.

\paragraph{Bursts and Jets as Amplified Quantum Fluctuations}

We now consider the evolution of quantum fluctuations, treated as
a test field riding on the collapsing condensate whose dynamics is
extracted from experiment. The initial state is defined by the
condition that $u=0$ for $t<0;$ we shall take it to be the
particle vacuum $\left| 0\right\rangle$, defined by $\psi _0\left(
x,0\right) \left| 0\right\rangle =0$ everywhere.

One can introduce a mode decomposition of the $\xi $ operator
based on the eigenfunctions of $\left[ H-E_{0}\right] H_{eff}$.
For short wavelengths $\lambda $, since $H\sim \lambda
^{-2}>>2u\Phi _{0}^{2}$, we expect these eigenfunctions will
approach the trap eigenmodes. The fact that particles in bursts
are seen to oscillate with the trap frequencies \cite{JILA01b}
also suggests that their dynamics is determined by the trap
Hamiltonian. Based on these observations we can assume a
homogeneous condensate $2u\Phi _{0}^{2}\sim \kappa ^{-1}a\omega
_{z}N_{0}\left( t\right) $, where $N_{0}\left( t\right) $ is the
instantaneous total number of particles in the condensate. In practice, $%
\kappa ^{-1}$ is a measure of the overlap between the condensate and the
excitation modes. Therefore, the approximation may be improved by adjusting $%
\kappa $ according to the range of modes where it will be applied.

Let $\bar N_0$ be the initial number of particles in the condensate, and $%
a_{cr}=\kappa /\bar N_0$ the corresponding critical scattering length. Trap
eigenfunctions $\psi _{\vec n}\left( r\right) $ are labelled by a string of
quantum numbers $\vec n=\left( n_z,n_x,n_y\right) .$ The eigenvalues of the
trap Hamiltonian are (with the zero energy already subtracted) $E_{\vec
n}=\omega _zn_z+\omega _\rho \left( n_x+n_y\right) $. There are two kinds of
modes, stable (oscillatory, or thawed) modes if $E_{\vec n}>\left( \frac
a{a_{cr}}\right) \omega _z,$ and unstable (growing, or frozen) modes if not.
In the former case we find that, although we assume vacuum initial
conditions, these modes do not remain empty. Up to $t_{collapse}$, when the
number of particles in the condensate is constant, the density

\begin{equation}
{\tilde n}\left( r,t\right) =\frac 18\left( \frac a{a_{cr}}\right) ^2\omega
_z^2\sum_{\vec n}\psi _{\vec n}^2\left( r\right) \frac{\sin ^2\omega _{\vec n}t}{\omega
_{\vec n}^2}\ \label{thawed}
\end{equation}
(where $\omega _{\vec{n}}=\sqrt{E_{\vec{n}}\left[ E_{\vec{n}}-\left( \frac{a}{%
a_{cr}}\right) \omega _{z}\right] }$) has a constant term and an oscillatory term. This
oscillatory term is responsible for the appearance of `\textbf{bursts}' of particles
oscillating within the trap observed in the Bose-Nova experiment \cite{JILA01b}. In the
WKB limit it describes a swarm of particles moving along classical trajectories in the
trap potential.

In the opposite case $E_{\vec n}\le \left( \frac a{a_{cr}}\right)
\omega _z,$ the formulae for the density is obtained by the
replacement of $\omega _{\vec n}$ in (\ref{thawed}) by $i\sigma
_{\vec n}$, thus $\omega _{\vec n}^{-1}\sin \omega _{\vec
n}t\rightarrow \sigma _{\vec n}^{-1}\sinh \sigma _{\vec n}t.$
Physically their difference is immense. In the first place, the
density is growing exponentially, but unlike the previous case,
there is no oscillatory component, and these particles do not
oscillate in the trap, in the sense described above. These modes
come alive at $\tau _{evolve}$ (as the scattering length is set
to zero), whence they become ordinary trap modes which oscillate
in the trap in the same way as the the burst modes . To the
observer, they appear as a new ejection of particles from the core
of the condensate, which makes up the so-called `\textbf{jets}'.
The sudden activation of a frozen mode (we are borrowing the
language and concept of cosmological structure formation) by
turning off the particle - particle interaction may be described
as a ``thaw''.

Observe that in this picture several conspicuous features of jets
become obvious. Jets may only appear if the turn - off time $\tau
_{evolve}$ is earlier than the formation time of the remnant. Once
the condensate becomes stable again, there are no more frozen
modes to thaw. On the other hand, jets will appear (as observed)
for $\tau _{evolve}<t_{collapse}$, when the condensate has not yet
shed any particles. Also jets must be less energetic than bursts,
since they are composed of lower modes.

Beyond $t_{collapse}$ the number of particles in the condensate, and
therefore the instantaneous frequency of the excited modes, becomes time
dependent. If we confine ourselves to the early stages of collapse we may
assume nevertheless that the condensate remains homogeneous. Shifting the
origin of time to $t_{collapse}$ for simplicity, we write $N_0\left(
t\right) =\bar N_0\mathrm{exp}\left( -t/\tau \right) $ (see Fig. 2).

\begin{figure}[h]
\includegraphics[height=3cm]{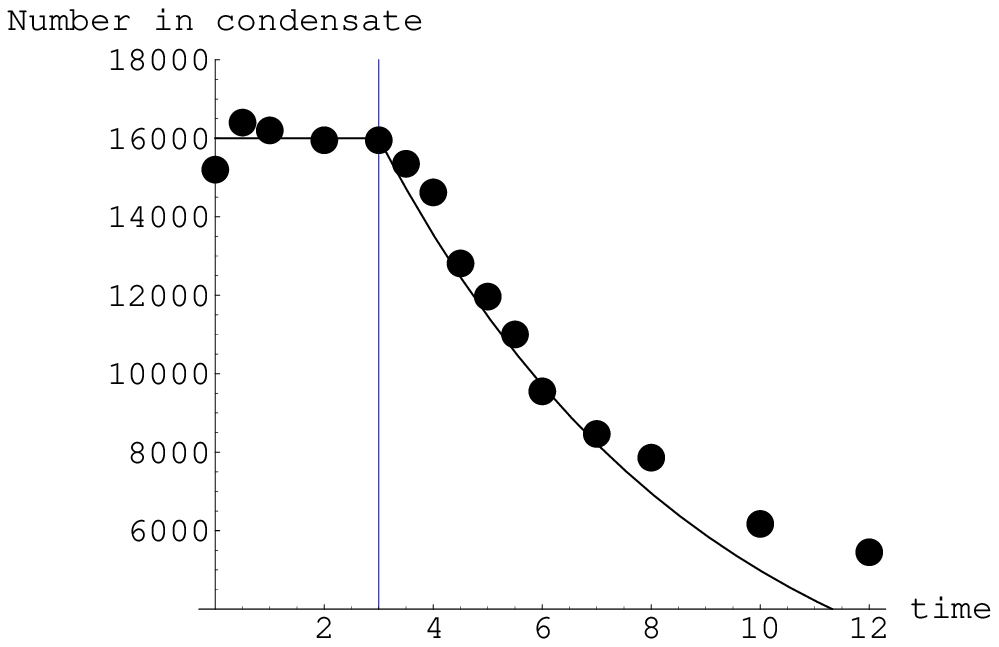}
\caption{{}}
\end{figure}

After expanding in trap eigenmodes we find the two kinds of behavior
described above. If $E_{\vec n}>\left( \frac{a\omega _z}{\bar a}\right) ,$
the mode is always oscillatory. If $E_{_{\vec n}}<\left( \frac{a\omega _z}{%
\bar a}\right) ,$ the mode is frozen at $t_{collapse},$ but thaws when $%
\mathrm{exp}\left( -t/\tau \right) \sim E_{\vec n}\bar a/a\omega _z$. During the frozen
period, the modes are amplified, but they only contribute to bursts after thawing. If the
evolution is interrupted while still frozen, they appear as a jet. We therefore conclude
that the number of particles $N_{jet}$ in a jet at time $\tau _{evolve}$ is essentially
the total number of particles in all frozen modes at that time. This is plotted in Fig 3,
for $\bar{N}_{0}=16,000,$
$\omega _{radial}=110$ Hz, $\omega _{axial}=42.7$ Hz, $a=36a_{0},$ and $%
\kappa =0.46$ , and compared to the corresponding results as reported in \cite{JILA01b}.

\begin{figure}[h]
\includegraphics[height=4cm]{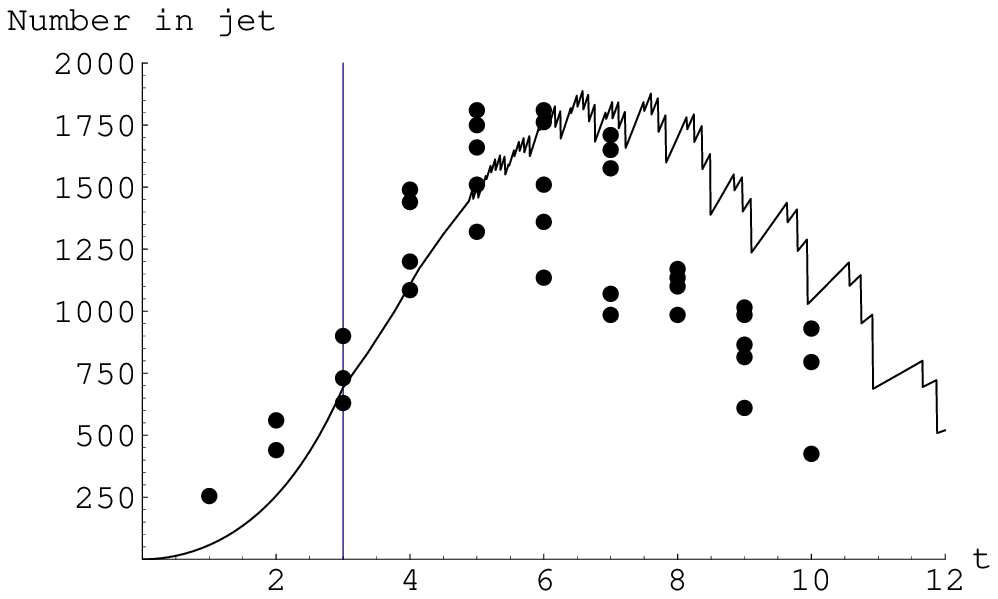}
\caption{}
\end{figure}

We see that the agreement is excellent at early times (up to
about $6$ms). For later times, this model overestimates the jet
population. This is due to the fact that, by not considering the
shrinking of the condensate, we are overestimating the overlap
between the condensate and the fluctuations, thus delaying the
thaw. It nevertheless reproduces the overall slope of particle
number with $\tau _{evolve},$. It should also be remembered that
we are computing the expected number of particles, but in the
highly squeezed state which results from the frozen period, the
fluctuations in particle number are comparable to the mean number
itself.

In this letter, we have presented a new viewpoint towards
understanding the salient features in the physics of controlled
collapse of a Bose-Einstein condensate described in the
experiment of \cite{JILA01b}, i.e, in terms of quantum vacuum
fluctuations parametrically amplified by the condensate dynamics.
Our way of thinking here is influenced by insights from the
quantum field theory of particle creation 
and structure formation 
in cosmological spacetimes as well as theories of spinodal
instability in phase transitions. 
One can conceivably  design experiments with BEC dynamics to test
out certain basic mechanisms and specific features of quantum
proceses in the early universe, thus opening a new venue for
performing `laboratory cosmology'. 


\noindent \textbf{Acknowledgement} We are obliged to E. Donley and
S. Kokkelmans for their prompt and informative responses to our
queries on some details of these experiments and for
communicating key unpublished data. EC also acknowledges
discussions with Eric Bolda. This research and EC's visits to UMD
are supported in part by a NSF grant PHY98-00967, a NIST grant
and by CONICET, UBA, Fundacion Antorchas and ANPCyT under grant
PICT99 03-05229.


\begin{thebibliography}{99}
\bibitem{JILA01b}  E. Donley et. al., Nature 412, 295 (2001); N. Claussen,
Ph. D. Thesis, U. of Colorado (2003).


\bibitem{JILA98}  J. Roberts et al., Phys. Rev. Lett. 81, 5109 (1998); S.
Cornish et al., Phys. Rev. Lett. 85, 1795 (2000).

\bibitem{JILA02a}  N. Claussen et al., Phys. Rev. Lett. 89, 10401 (2002); E.
Donley et al., Nature 417, 529 (2002).

\bibitem{KGB02}  S. Kokkelmans and M. Holland, Phys. Rev. Lett. 89, 180401
(2002); M. Mackie, K. Suominen and J. Javainen, Phys. Rev. Lett.
89, 180403 (2002); cond-mat/0209083; M. Mackie et al.,
physics/0210131.

\bibitem{JILA03}  N. Claussen et al., cond-mat/0302195; S.
Kokkelmans, private communication.

\bibitem{CHBosenova}  E. Calzetta and B. L. Hu, cond-mat/0207289 v2.

\bibitem{Savage}  C. Savage, N. Robins and J. Hope, cond-mat/0207308.

\bibitem{DS02} R. Duine and H. Stoof, cond-mat/0211514.

\bibitem{KM01}  Yu. Kagan and L. Maksimov, Phys. Rev. A64, 53610 (2001).

\bibitem{Y02}  V. A. Yurovsky, Phys. Rev. A65, 33605 (2002). V. A. Yurovsky A.
Ben-Reuven, cond-mat/0205267

\bibitem{M99}   S. A. Morgan et al., Phys. Rev. A 57, 3818 (1998).

\bibitem{GTT01}  A. Gammal, T. Frederico and L. Tomio, Phys. Rev. A64, 55602
(2001).

\bibitem{TBJ00}  M. Trippenbach, Y. Band and P. Julienne, Phys. Rev. A62,
23608 (2000).

\bibitem{SU03}  H. Saito and M. Ueda, cond-mat/0305242.

\end{thebibliography}
\end{document}